\documentclass[aps,prd,twocolumn,superscriptaddress,nofootinbib,groupedaddress,preprintnumbers]{revtex4}
\pdfoutput=1

\usepackage{slashed}
\usepackage{color, verbatim}
\usepackage{latexsym}
\usepackage{amssymb}
\usepackage{amsmath}
\usepackage{graphicx}\graphicspath{{figs/}}
\usepackage[dvipsnames]{xcolor}
\usepackage{adjustbox}
\usepackage{easybmat}
\usepackage{bbm}
\usepackage{bm}
\usepackage{adjustbox}
\usepackage{booktabs}
\usepackage{multirow}
\usepackage{rotating}
\usepackage{soul}
\usepackage{hyperref}
\usepackage{mathtools}
\usepackage{wasysym}

\allowdisplaybreaks

\newcommand{\ifb}{\text{ fb}^{-1}}
\newcommand{\iab}{\text{ ab}^{-1}}

\newcommand{\gev}{\,\text{GeV}}
\newcommand{\tev}{\,\text{TeV}}

\newcommand{\mOSSF}{M_\text{OSSF}^{20}}

\newcommand{\etmiss}{E_T^\text{miss}}

\begin{document}
\preprint{IPPP/22/16}
\title{New Higgs decays to axion-like particles}

\author{Anke Biek\"otter}
\email{anke.biekoetter@durham.ac.uk}
\affiliation{Institute for Particle Physics Phenomenology, Department of Physics, Durham University, South Road, Durham DH1 3LE, United Kingdom}

\author{Mikael Chala}
\affiliation{Departamento de F\'isica Te\'orica y del Cosmos, Universidad de Granada, E--18071 Granada, Spain}
\email{mikael.chala@ugr.es}
\author{Michael Spannowsky}
\email{michael.spannowsky@durham.ac.uk}
\affiliation{Institute for Particle Physics Phenomenology, Department of Physics, Durham University, South Road, Durham DH1 3LE, United Kingdom}

\begin{abstract} 
We investigate the interactions of a light scalar with the Higgs boson and second-generation fermions, which trigger new rare decays of the Higgs boson into $4\mu$, $2\mu 2\gamma$, $6\mu$ and $4\mu 2j$. 
We recast current LHC searches to constrain these decays and develop new collider analyses for those channels which are only poorly tested by existing studies. 
With the currently collected data we can probe branching ratios as small as $1.5 \times10^{-5}$, $8.7 \times10^{-5}$, $5.7 \times10^{-8}$ and $1.6 \times10^{-7}$, respectively. For the High-Luminosity LHC run, considered here to involve 3 ab$^{-1}$ of integrated luminosity, these numbers go down to $1.3 \times10^{-5}$, $2.0 \times10^{-6}$, $3.0 \times10^{-9}$ and $5.4 \times10^{-9}$, respectively. 
 We also comment on other channels that remain still unexplored. 
\end{abstract}

\maketitle

\newpage

\section{Introduction}
\label{sec:intro} 
Light neutral (pseudo-)scalar particles are ubiquitous in models of new physics, including theories with a spontaneously broken global symmetry such as the Peccei-Quinn solution to the strong CP problem~\cite{Peccei:1977hh,Peccei:1977ur,Weinberg:1977ma,Wilczek:1977pj}, composite Higgs models~\cite{Gripaios:2009pe,Gripaios:2016mmi,Chala:2017sjk} and others~\cite{Wilczek:1982rv,Chikashige:1980ui}. Moreover, they appear in different explanations for dark matter~\cite{Preskill:1982cy,Abbott:1982af,Dine:1982ah} as well as for the flavour problem~\cite{Davidson:1981zd,Wilczek:1982rv,Ema:2016ops,Calibbi:2016hwq} and the hierarchy problem~\cite{Graham:2015cka}.
With a little abuse of language, we will hereafter refer to these light scalars as axion-like particles (ALPs), irrespective of whether or not they fulfil a shift symmetry $a\to a+a_0$. In-depth analyses, including quantum effects, of the possible interactions between these ALPs and the SM fields, have been recently presented in Refs.~\cite{Chala:2020wvs,Bauer:2020jbp,Galda:2021hbr,Bonilla:2021ufe}.

Many of those couplings are constrained by data collected at low-energy and flavour facilities~\cite{Povey:2010hs,ADMX:2010ubl,Betz:2013dza,Bjorken:2009mm,Andreas:2012mt,Essig:2013lka,Harland-Lang:2019zur,Gavela:2019wzg,Calibbi:2020jvd,MartinCamalich:2020dfe,Bauer:2021mvw,Carmona:2021seb,Jerhot:2022chi} 
colliders including LEP \cite{Jaeckel:2015jla,Mimasu:2014nea,Bauer:2017ris,Craig:2018kne} and the  LHC~\cite{Jaeckel:2012yz,Knapen:2016moh,Brivio:2017ije,Bauer:2017ris,Bauer:2017nlg,Alonso-Alvarez:2018irt,Craig:2018kne,Ebadi:2019gij,Gavela:2019cmq,Coelho:2020saz,Haghighat:2020nuh,Goncalves:2020bqi,Carmona:2022jid} or in astrophysical events~\cite{Raffelt:2006cw,CAST:2011rjr,Lee:2018lcj,Chang:2018rso,Jaeckel:2019xpa,Alonso-Alvarez:2019ssa,Ertas:2020xcc}.
Some couplings have been shown to be only testable in Higgs physics, particularly through yet unexplored decays of the Higgs boson. Recent analyses in this line include flavour-violating ALPs~\cite{Evans:2019xer,Davoudiasl:2021haa} and a search for Higgs decays to ALPs with subsequent decay to photons and jets explored in~\cite{Alves:2021puo}; see also Ref.~\cite{Cepeda:2021rql} for an extensive review on the topic.

In this paper, despite continuing this research avenue, we focus on flavour-conserving Higgs decays in which not all final fermion pairs reconstruct narrow masses as our targeted signals ensue from Higgs-fermion-ALP contact interactions. The article is organised as follows. In section~\ref{sec:theory} we introduce the relevant ALP interactions with Standard Model (SM) particles; we discuss their possible origin and estimate the resulting Higgs branching ratios into the final states of interest.  In section~\ref{sec:analysis} we present a thorough analysis of the potential of current experimental data.  We recast sensitive existing searches, develop new search strategies and set limits on the ALP interactions with the SM.  We conclude in section~\ref{sec:conclusions}.  We dedicate Appendix~\ref{app:4mu2j} to limitations of other potential analyses not considered in the main text.

\section{Theory and experiment} 
\label{sec:theory}
From an agnostic point of view, and without exact knowledge of the nature of the electroweak symmetry breaking (EWSB), namely whether it is linearly or non-linearly realised, the interactions between the ALP and the SM particles must be described by an effective-field theory, the Lagrangian of which reads:
\begin{align}
 L &=   \frac{1}{2} \partial_\mu a \partial^\mu a - \frac{m_a^2}{2} a^2 \nonumber\\
    &+c_{a A} \frac{a}{f_a}  A_{\mu\nu} \tilde{A}^{\mu \nu} 
    +c_{a \psi} a  \overline{\psi}\psi 
   + c_1^\psi \frac{a}{f_a} h \overline{\psi}\psi + c_2^\psi \frac{a^2}{f_a^2} h  \overline{\psi}\psi \,
   \label{eq:lagrangian}
\end{align}
plus other terms that are either suppressed by further powers $f_a$ or that involve other gauge bosons, which we are not interested in.
Note that $h$ represents the physical Higgs boson, $A$ stands for the photon, and $\psi$ can be any fermion; we ignore family indices as we assume flavour conservation.
We only consider second-generation fermions in the following, since their couplings, particularly those involving leptons (muons), can almost exclusively be tested in Higgs decays at the LHC. There are currently no other facilities in which the intervening particles can collide at sufficiently high energies.

\begin{figure*}[t]
 \includegraphics[width=0.24\textwidth]{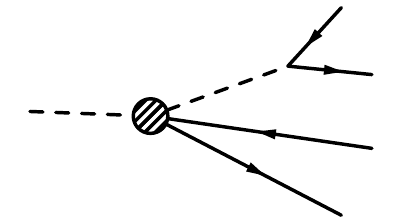}
 \includegraphics[width=0.24\textwidth]{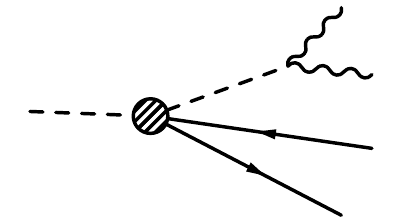}
 \includegraphics[width=0.24\textwidth]{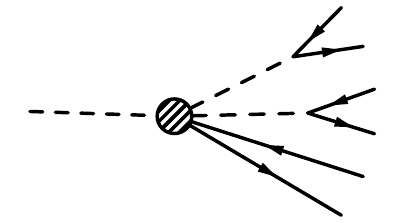}
 \includegraphics[width=0.24\textwidth]{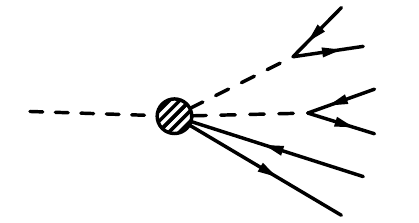}
 \caption{Diagrams representing the Higgs decays of interest: $h\to 4\mu$ (left), $h\to 2\mu2\gamma$ (center-left), $h\to 6\mu$ (center-right), $h\to 4\mu 2j$ (right).}
 \label{fig:feyn_diags}
\end{figure*}

The $h a \bar{\psi} \psi$ interactions can be sizeable even if $a \bar{\psi} \psi$ couplings are small or absent. 
If the EWSB is linearly realised, then Eq.~\eqref{eq:lagrangian} arises from a more fundamental Lagrangian involving the full Higgs doublet $H=[G^+,(h+i G^0)/\sqrt{2}]^T$ upon EWSB; $h\to h+v$ with $v\sim 246$ GeV. 
Even in this case, $c_{1}^\psi$ is not necessarily related to $c_{a\psi}$, being  therefore, in general, unconstrained. To see this, consider the following Lagrangian:
\begin{align}
 L =\frac{a}{f_a} \overline{\psi_L} H \psi_R \left(c_1 + c_2 \frac{|H|^2}{f_a^2} \right)+\text{h.c.}
\end{align}
One can, without loss of generality, define the new coupling $c_1' = c_1+c_2 v^2/(2f_a^2)$, from where we obtain the following Lagrangian after EWSB:
\begin{align}
 L = c_1' \frac{v}{\sqrt{2} f_a} a  \overline{\psi}\psi +\left(c_1' +  c_2 \frac{v^2}{f_a^2} \right) \frac{a}{\sqrt{2} f_a} h \overline{\psi}\psi\,.
\end{align}
Since $c_1$ was completely arbitrary, so is $c_1'$. The fact that the coupling $c_1'$, which induces decays of the ALP to fermions, is small as required by experimental data~\cite{Bauer:2017ris,Liu:2018xkx}, implies some tuning between two couplings. In particular, it implies that $c_1$ is smaller than naively anticipated based on $v/f_a$ power counting alone. This possibility, however, must also be considered and so it is possible to have sizeable $ha \overline{\psi}\psi$ couplings in the absence of large $a \overline{\psi}\psi$ interactions.

The different ways in which the ALP $a$ can decay remain largely unconstrained. Thus, $a\to\mu^+\mu^-$ can dominate if, for example, a $\mathbb{Z}_2$ symmetry $a\to -a$ is only broken in the muon sector while being exact in the quark sector. This forbids $h\to aqq$, but allows $h\to a \mu\mu$.
Likewise, it makes $c_2^\psi$ vanish for neither leptons nor quarks, so both $h\to aaqq$ and $h\to aa\mu\mu$ remain open in principle. However, the fact that both can coexist is of little relevance for our analysis, as the SM dominates the Higgs width; $\Gamma_h\sim 4$~MeV~\cite{LHCHiggsCrossSectionWorkingGroup:2016ypw}.
On the other hand, $a\to\gamma\gamma$ can be sizeable (or even the dominating decay) if the $\mathbb{Z}_2$ symmetry $a\to -a$ and $h\to -h$ is respected in good approximation by all fermions and provided $c_{aA}$ is non-zero (Note that even if $c_{aA}\sim \alpha/(4\pi)$, as suggested by naive power counting, the decay $a\to\gamma\gamma$ is prompt.).

In light of this discussion, we focus on the Higgs decays to ALPs with subsequent decays of the ALP to muons or photons in this article.
In particular, we investigate the following Higgs decay modes which are shown in Fig.~\ref{fig:feyn_diags}: (i) $h\to a \mu\mu, a\to\mu\mu$; (ii) $h\to a \mu\mu, a\to\gamma\gamma$; (iii) $h\to aa\mu\mu, a\to\mu\mu$; (iv) $h\to aa qq, a\to\mu\mu$. We restrict the ALP mass range to $m_a\in [10,50]$ in order to stay well below the $Z$~boson peak for di-muon resonances and to ensure a large enough angular separation between the photons in low-mass di-photon resonances. 
Apart from Higgs decays, the processes (i)--(iii) can only be tested at muon colliders in the considered ALP mass range. However, process (iv) can also be tested at the LHC through $p p \to h a a, \, a \to \mu \mu$ and we will indeed see in section~\ref{sec:analysis_4mu2q} that this channel provides more substantial limits on the corresponding Wilson coefficient.

To test these Higgs decays, we recast current LHC analyses at $\sqrt{s}=13 \tev$, where possible. However, we refrain from using $7\,\tev$ or $8\,\tev$ analyses due to the lower Higgs production cross-section and luminosity.  
We have performed a thorough and systematic study of experimental searches in final states involving at least two muons to find suitable analyses. 
We notice that most beyond SM searches are of little relevance.  SUSY searches typically enforce substantial cuts on the missing energy $\etmiss$ absent within our signals, whereas
heavy resonance searches, on the other hand, often require large transverse momenta of the final-state particles, while the decay products of a four- or six-body decay of a $125\gev$ resonance are rarely very energetic.

Before discussing search strategies for our exotic Higgs decays, let us estimate the corresponding partial decay widths. This allows us to project the number of events for current and future runs of the LHC and to establish the signal acceptance needed for an analysis to be sensitive to our ALP signals. 

We concentrate first on the four-body decays of the Higgs, $h\to 4\mu$ and $h \to 2\mu 2\gamma$.  The partial width for
the Higgs decay
to an ALP and two fermions can be estimated to be:
\begin{equation}
 \Gamma(h\to a\overline{\psi}\psi)\sim \frac{(c^\psi_1)^2}{768\,\pi^3 f_a^2} m_h^3 \,.
\end{equation}
For $\mathcal{O}(1)$ couplings at $f_a=1$ TeV, the corresponding branching ratio is about $10^{-3}$.
Taking into account the Higgs production cross section at $\sqrt{s}=13$ TeV which is approximately $48$ pb~\cite{Anastasiou:2016cez} and the planned integrated luminosity of $\mathcal{L}=3$ ab$^{-1}$ we expect 
$\sim 10^{6}$
$h \to a \overline{\psi}\psi$ events at the High-Luminosity LHC (HL-LHC). 
Optimistically assuming that at least 10 signal events are needed to distinguish the signal from background fluctuations at the $2\sigma$ level (as these two channels are never background-free), it can be straightforwardly concluded that an analysis probing the two Higgs decays requires at least a signal acceptance of $10^{-5}$. 

Let us now turn our attention to the six-body Higgs decays in the channels $h\to 6\mu$ and $h\to 4\mu 2j$.
These processes result from the decay of a Higgs into two ALPs and two fermions with an approximate width:
\begin{equation}
 \Gamma(h\to aa\overline{\psi}\psi)\sim \frac{(c^\psi_1)^2}{73728\,\pi^5 f_a^4}m_h^5\,.
\end{equation}
The corresponding branching ratio is of about $10^{-7}$ for $\mathcal{O}(1)$ couplings at $f_a = 1 \tev$ and we expect only 
$\sim 50$ $h\to a a \overline{\psi}\psi$ events to be produced at the HL-LHC.
Even background-free experimental searches can thus only be sensitive to $\mathcal{O}(1)$ couplings triggering these decays provided their signal acceptance is at least $10 \%$.  

\section{Analysis}
\label{sec:analysis}
To set limits on the ALP interactions in our model, we compare the predicted fiducial cross-sections with current LHC data. 
We have implemented the interactions of the Lagrangian in Eq.~\eqref{eq:lagrangian} in FeynRules~\cite{Degrande:2011ua}. Events have been generated using \texttt{MadGraph-v2.7.3}~\cite{Alwall:2014hca}, showered with \texttt{Pythia8}~\cite{Sjostrand:2007gs} and analysed using \texttt{Rivet-v2.7.0}~\cite{Buckley:2010ar}.

Unless otherwise stated, all signal and background samples have been generated at leading order. For gluon fusion Higgs production, we include a $k$~factor to normalize the cross-section to its N$^3$LO QCD$+$NLO EW prediction of $48.58$~pb~\cite{Anastasiou:2016cez}.

\subsection{Four muon final state}
\label{sec:analysis_4mu}

\begin{figure}
    \centering
     \includegraphics[width=.45\textwidth]{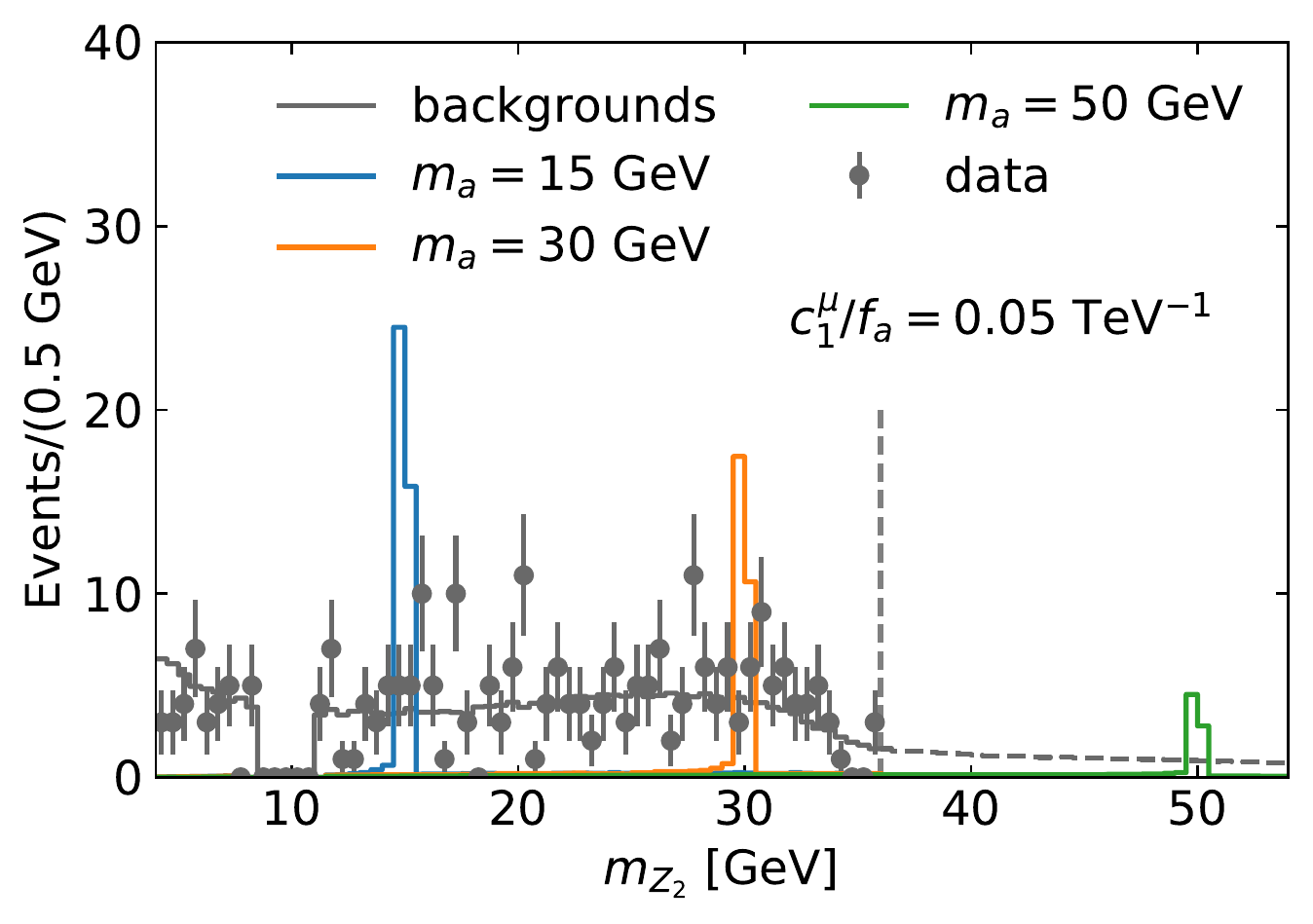}
     \includegraphics[width=.45\textwidth]{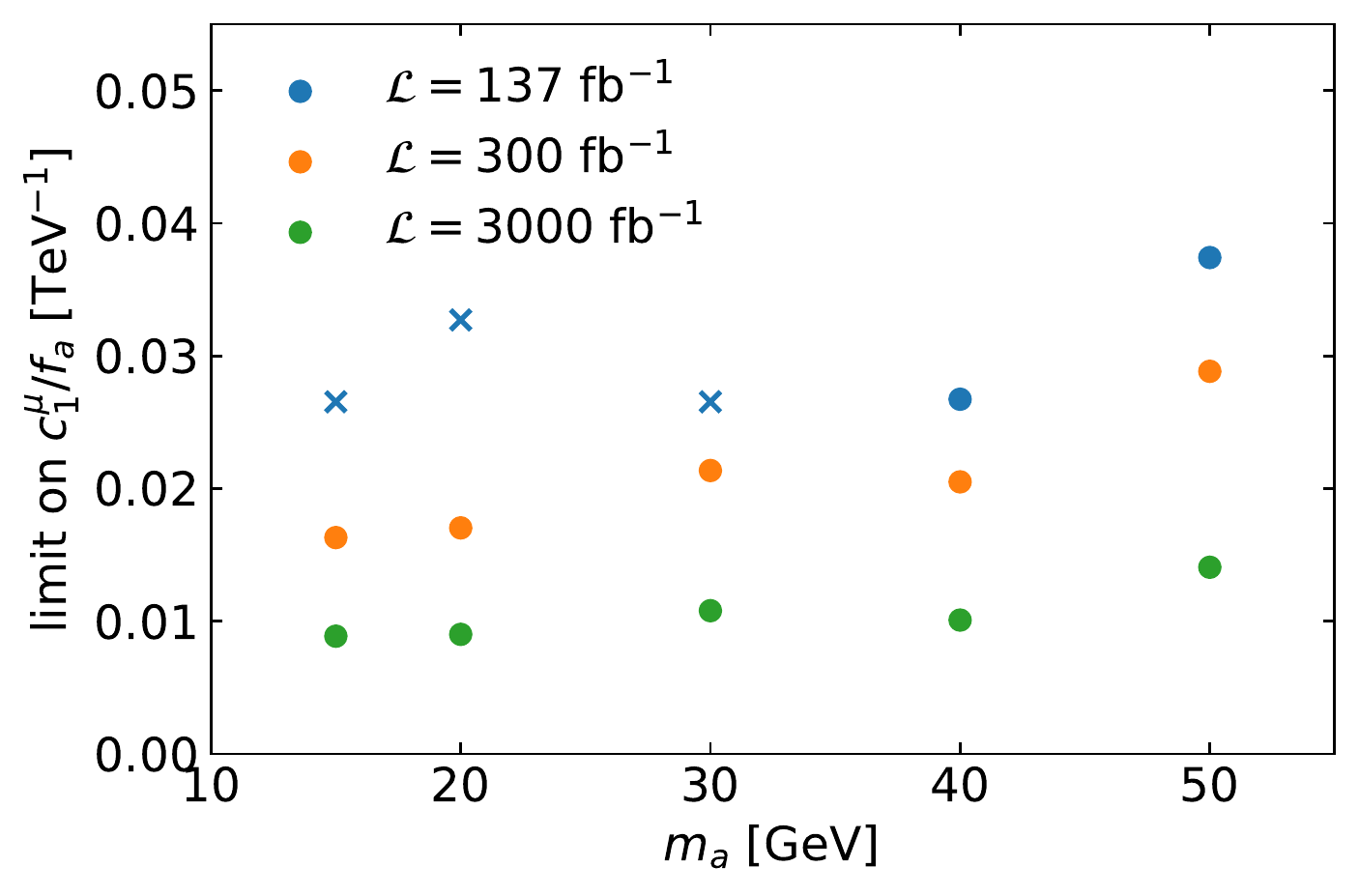}
	\caption{Top: $m_{Z_2}$ distribution in the $h \to 4 \mu$ analysis for data,  background and signals at different masses ($\mathcal{L} = 137\ifb$).  The gray dashed line at $m_{Z_2} = 36 \gev$ represents the upper limit of the search region considered in the CMS analysis~\cite{CMS:2021pcy};  see text for details. 
	Bottom: Upper $95\%$ CLs limits on $c_1^\mu/f_a$ at different luminosities. Limits marked with an ``x'' are observed limits, other limits are expected limits. }
	\label{fig:4mu_mZ2}
\end{figure}

To constrain the Wilson coefficient $c_1^\mu$, we analyse contributions to Higgs decays in the four-lepton final state via $h\to a \mu\mu \to 4 \mu$ as represented by the Feynman diagram in the left panel of Fig.~\ref{fig:feyn_diags}.

Several LHC searches analyse final states with four muons, rendering them potentially sensitive to our 
$h\to 4\mu$ signal. 
However, most SM analyses focus on different invariant mass ranges for the four-lepton system, see e.g. ~Ref.~\cite{CMS:2018yxg} for a $Z\to 4\mu$ analysis, and most exotic/SUSY searches focus on high-energy events with large total invariant or effective masses, see Refs.~\cite{ATLAS:2018rns,CMS:2020zjg}.

An analysis that targets explicitly our signal properties is the CMS search for low-mass di-lepton resonances in Higgs boson decays to four leptons~\cite{CMS:2021pcy}.
The analysis requires two opposite-sign same-flavour (OSSF) lepton pairs with a four-lepton invariant mass close to the Higgs mass:
\begin{equation}
\begin{split}
p_T^{\ell} > 20, \, 15, \, 5, &\, 5 \gev \, , \quad 
m_{4\ell} \in [118, \, 130] \gev \, , \\
m_{\ell^+\ell^-} > 4& \gev ,\quad
 m_{\ell^+\ell^-} \not\in [8.0, \, 11.5] \gev  \\
m_{Z_1} > 40 &\gev , \quad
\Delta R (\ell_i , \ell_j ) > 0.02 \, .
\end{split}
\label{eq:cuts_4mu}
\end{equation}
For the ZX($\mu\mu$) search region of the analysis, the leptons are paired such that $Z_1$ is the OSSF lepton pair closest to the $Z$~boson mass, i.e.\ for which $|m_{Z_1} - m_Z|$ is minimal. The second OSSF lepton pair $Z_2$ is required to consist of two muons. 
Events with an OSSF lepton pair in the invariant mass range $m_{\ell^+\ell^-} \in [8.0, \, 11.5] \gev$ are excluded to suppress the background from $\Upsilon (b b \bar )$ decays. We have implemented the cuts in Eq.~\eqref{eq:cuts_4mu} in \texttt{Rivet}. 

We present the mass distribution of the second OSSF lepton pair~$m_{Z_2}$ in the top panel of Fig.~\ref{fig:4mu_mZ2}.  
A Higgs decay via the $h \to a \mu \mu \to 4 \mu$ channel leads to a resonance at $m_{Z_2} \sim m_a$ (in the considered $m_a$ mass range).
Therefore, we can use this distribution to set limits on the Wilson coefficient $c_1^\mu$. 
The dominant backgrounds for this analysis are $h \to ZZ^*$ and $ZZ$ production, and we take their prediction directly from Fig.~2a of Ref.~\cite{CMS:2021pcy} along with a $10\%$ systematic uncertainty on the total per-bin background prediction.
The $h \to ZZ^* \to 4 \ell$ background prediction was used to validate our implementation of the CMS analysis in \texttt{Rivet}. We agree with the number of events in each of the 64 bins at the $10\%$ level after including an experimental efficiency of $74 \%$. 

The corresponding observed (expected) $95\,\%$~CLs limits~\cite{Read:2002hq} at $\mathcal{L}=137 \ifb$ are:
\begin{equation}
 \frac{c_1^\mu}{f_a} <
\begin{aligned}
0.027 \,(0.021)/\text{TeV} \quad  \text{for }m_s = 15\gev \\  
0.027 \,(0.025)/\text{TeV} \quad \text{for }m_s = 30\gev \\
\end{aligned}
  \,  ,
\end{equation}
assuming that all ALPs decay to muons only. 

While the CMS search only considers the region $m_{Z_2}\leq 36\gev$, we can in principle extend the search region to higher $Z_2$ masses.  The background 
at these masses is dominated by SM Higgs decays and we rely on our own background prediction in this region. 
The expected $95\,\%$~CLs limit for a $m_a=50 \gev$ ALP and $137 
\ifb$ of data are:
\begin{equation}
 \frac{c_1^\mu}{f_a} <0.037/\text{TeV} \quad \text{for }m_s = 50\gev \\
  \,  .
\end{equation}
In the bottom panel of Fig.~\ref{fig:4mu_mZ2}, we show the limits on $c_1^\mu/f_a$ for different ALP masses.  
We present the limits for the current luminosity $137\ifb$ 
as well as for the planned luminosities after LHC Run~III, $300\ifb$, and the 
HL-LHC, $3 \iab$.
For higher ALP masses, the limits slightly decrease due to the larger phase-space suppression; also, compare the top panel of Fig.~\ref{fig:4mu_mZ2}. However, the reduced production cross-section is only partially compensated by a smaller background contribution and a higher signal acceptance as a result of the muons carrying on average higher transverse momenta.

The sensitivity of the current CMS analysis to our ALP signal could be further improved by requiring that all four leptons, and not only the ones forming $Z_2$, are muons. This would decrease the background contribution by an approximate factor of $1/2$ and tighten the expected limits by around $10\,\%$.

\subsection{Two muon two photon final state}
\label{sec:analysis_2mu2a}

\begin{figure}[thb]
		\centering
		\includegraphics[width=.45\textwidth]{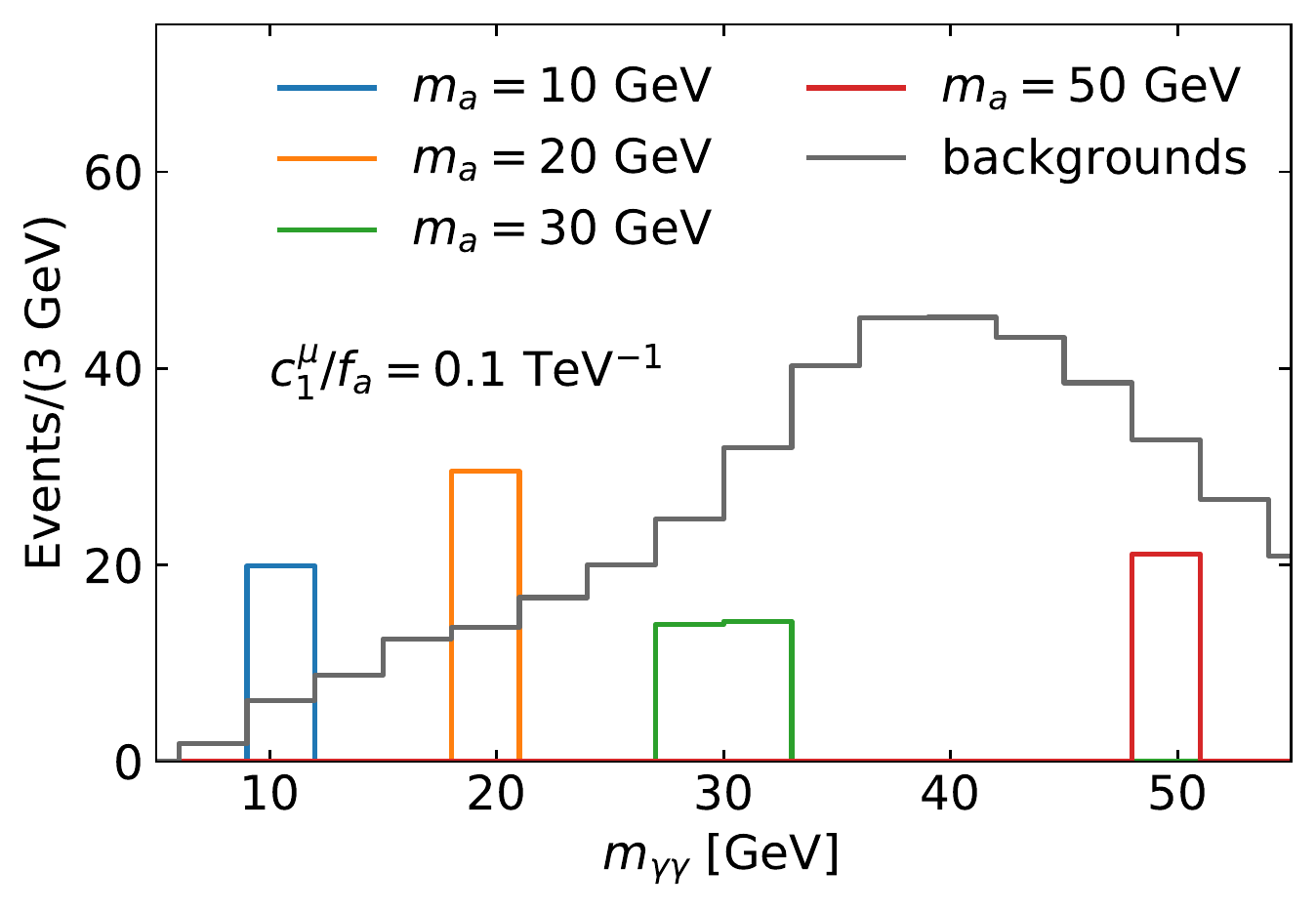} 
		\includegraphics[width=.45\textwidth]{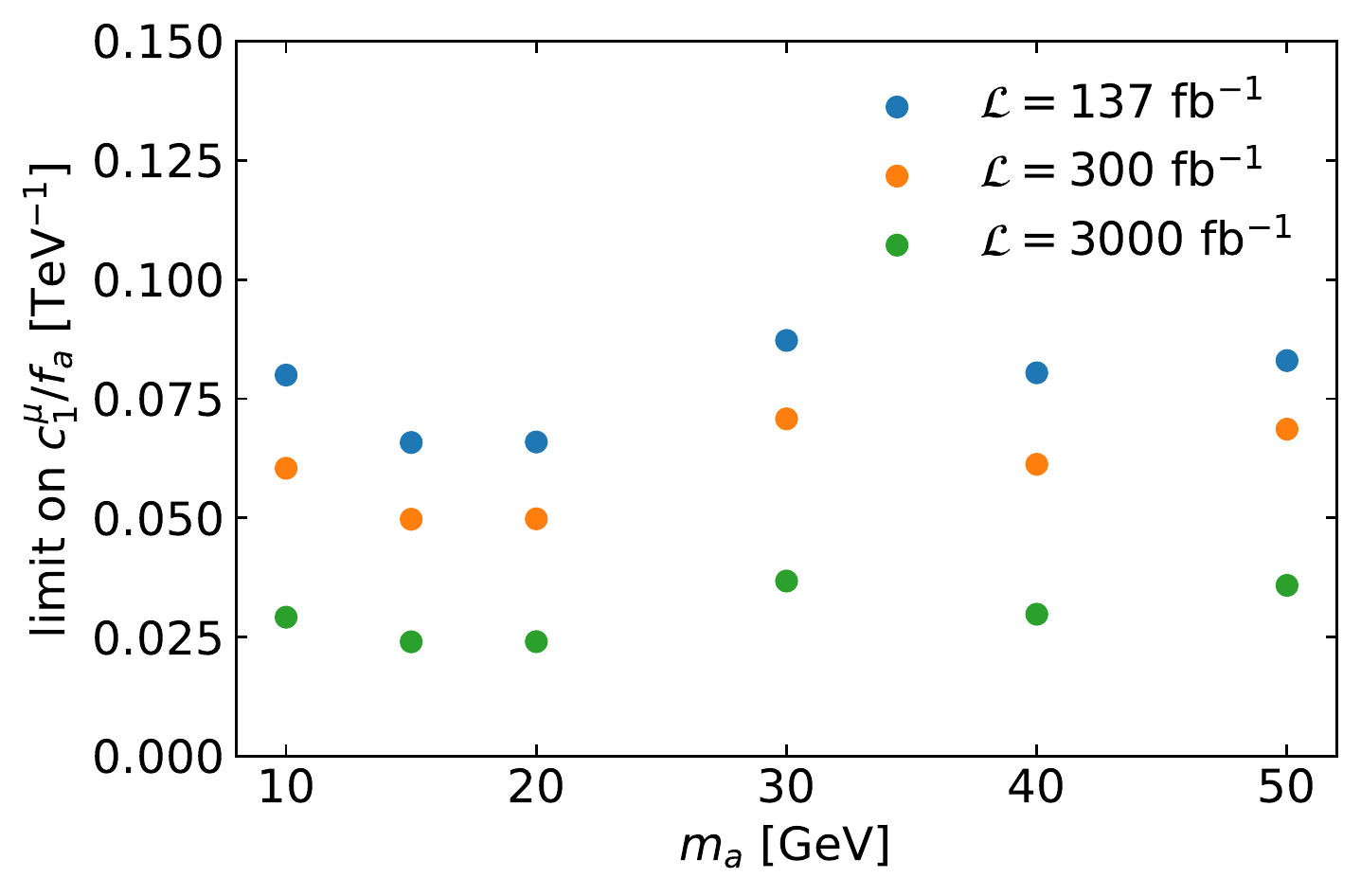} 
		\caption{Top: $m_{\gamma\gamma}$ distribution in the $h\to a \mu\mu \to 2\mu 2\gamma$ analysis for signal and backgrounds at $\mathcal{L} = 137\ifb$.  Bottom: Upper $95\%$ CLs limits on $c_1^\mu/f_a$ for different luminosities.  }
		\label{fig:2mu2gamma}
\end{figure}

We now consider the possibility that the ALP is produced in a Higgs decay in association with two muons and that it decays into two photons.  The corresponding Feynman diagram is shown in the second panel of Fig.~\ref{fig:feyn_diags}.

The most promising LHC analysis to constrain our $2\mu 2\gamma$ ALP signal is the CMS search targeting events with a leptonically decaying $Z$ boson and two photons~\cite{CMS:2021jji}
\footnote{Searches for Higgs decays in the $h \to Z \gamma \to \ell \ell \gamma$ channel also offer some sensitivity to our $h \to 2\mu 2\gamma$ signal.  However,  due to the fact these analysis target events with a single photon, we cannot use $h \to Z \gamma $ searches to constrain the ALP mass which shows up as a resonance in the $m_{\gamma\gamma}$ spectrum.  
Moreover,  the ATLAS search~\cite{ATLAS:2020qcv} uses a boosted decision tree (BDT) which makes it impossible to recast without detailed information on the BDT.   Moreover, the CMS search~\cite{CMS:2018myz} is currently only published for a low luminosity ($\mathcal{L} = 35.9 \ifb$).  
}.
However, its cut on the transverse momentum of the two photons, $p_T^\gamma > 20 \gev$,  is only rarely passed for the photons resulting from the decay of a low-mass ALP $a \to \gamma \gamma$.
To increase the sensitivity on Higgs decays to ALPs via the $h \to a 2 \mu$ channel,  we design our own analysis based on Ref.~\cite{CMS:2021jji}. 
We relax the cuts  $m_{\ell\ell} > 55 $~GeV and $p_T^\gamma > 20\gev$ in the experimental reference to make the analysis more inclusive:
\begin{equation}
\begin{split}
p_T^\mu >  30, \,  &15 \gev \, , \quad 
|\eta^\mu| < 2.4 \gev \, , \quad 
m_{\mu \mu} > 12 \gev \\
|\eta^\gamma| &< 2.5 \gev \, (\text{excl. } 1.44 < |\eta^\gamma| < 1.57), \\
&p_T^\gamma  > 15,  \gev \, , \quad
 \Delta R ( \gamma , \, \mu/\gamma ) > 0.4
\, .
\end{split}
\end{equation}
To select the Higgs decay region we further require $m_{\ell \ell \gamma \gamma} \in [110, 140] \gev$.
 
To set limits, we consider the invariant mass distribution of the two photons~$m_{\gamma \gamma}$, where the ALP signal will produce a resonance at its mass; compare the top panel of Fig.~\ref{fig:2mu2gamma}. 
 
The dominant backgrounds are $Z\gamma \gamma$ and $Z+$jets production, where the jets fake photons. 
To estimate the size of these backgrounds, we have generated $ \mu^+ \mu^- \gamma \gamma$ production at NLO in QCD using MadGraph. 
To account for the $Z+$jets background, we have multiplied the generated background distribution by a factor $1.7$. This factor is based on the  $Z+\,$jets production background contribution in $Z\gamma\gamma$ production~\cite{CMS:2021jji},  which is $\sim 62\, \%$ the size of the signal. 
We conservatively assume a $25\%$ systematic uncertainty on the background prediction based on the uncertainty of $Z+\,$jets production background contribution; see Ref.~\cite{CMS:2021jji}.

The $95\,\%$ CLs limits resulting from the $m_{\gamma \gamma}$ distribution for $137 \ifb$ are:
\begin{equation}
 \frac{c_1^\mu}{f_a} <
 \begin{aligned}
0.066/\text{TeV}^2 \quad  \text{for }m_s = 15\gev \\  
0.083/\text{TeV}^2 \quad \text{for }m_s = 50\gev 
 \end{aligned}
  \,  .
\end{equation}
In the bottom panel of Fig.~\ref{fig:2mu2gamma}, we show the limits on $c_1^\mu/f_a$ for different ALP masses and luminosities. 
The limits 
depend only very weakly on the ALP mass $m_a$ as a result of three competing effects:
First, the Higgs branching ratio into ALPs decreases with $m_a$ due to the phase-space suppression. Second, the signal acceptance increases with $m_a$. And third, the background distribution is not flat but peaks around $40 \gev$;
see the upper panel of Fig.~\ref{fig:2mu2gamma}.

\subsection{Six muon final state}
\label{sec:analysis_6mu}
\begin{figure}[thb]
		\centering
		\includegraphics[width=.45\textwidth]{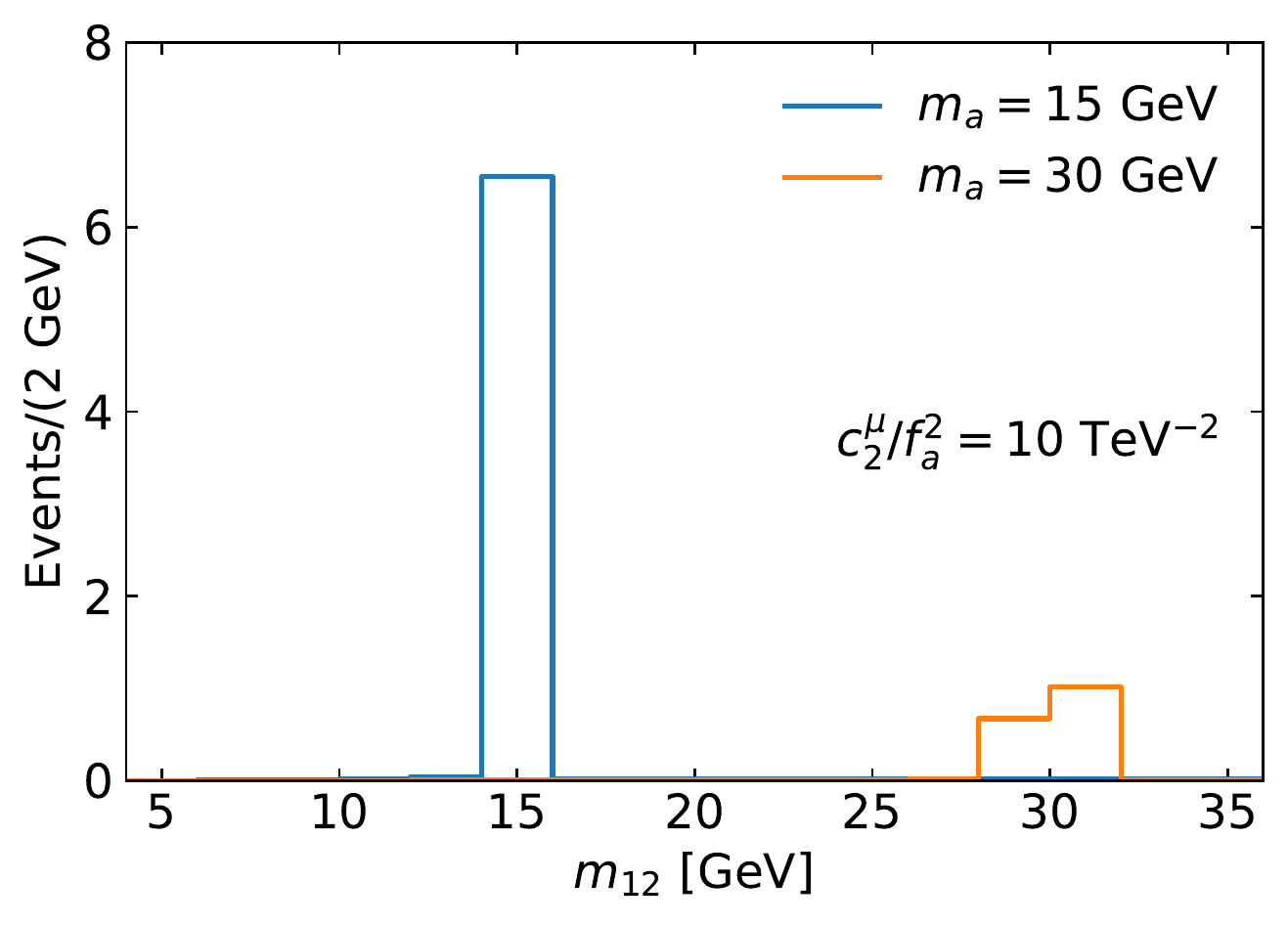} 
		\includegraphics[width=.45\textwidth]{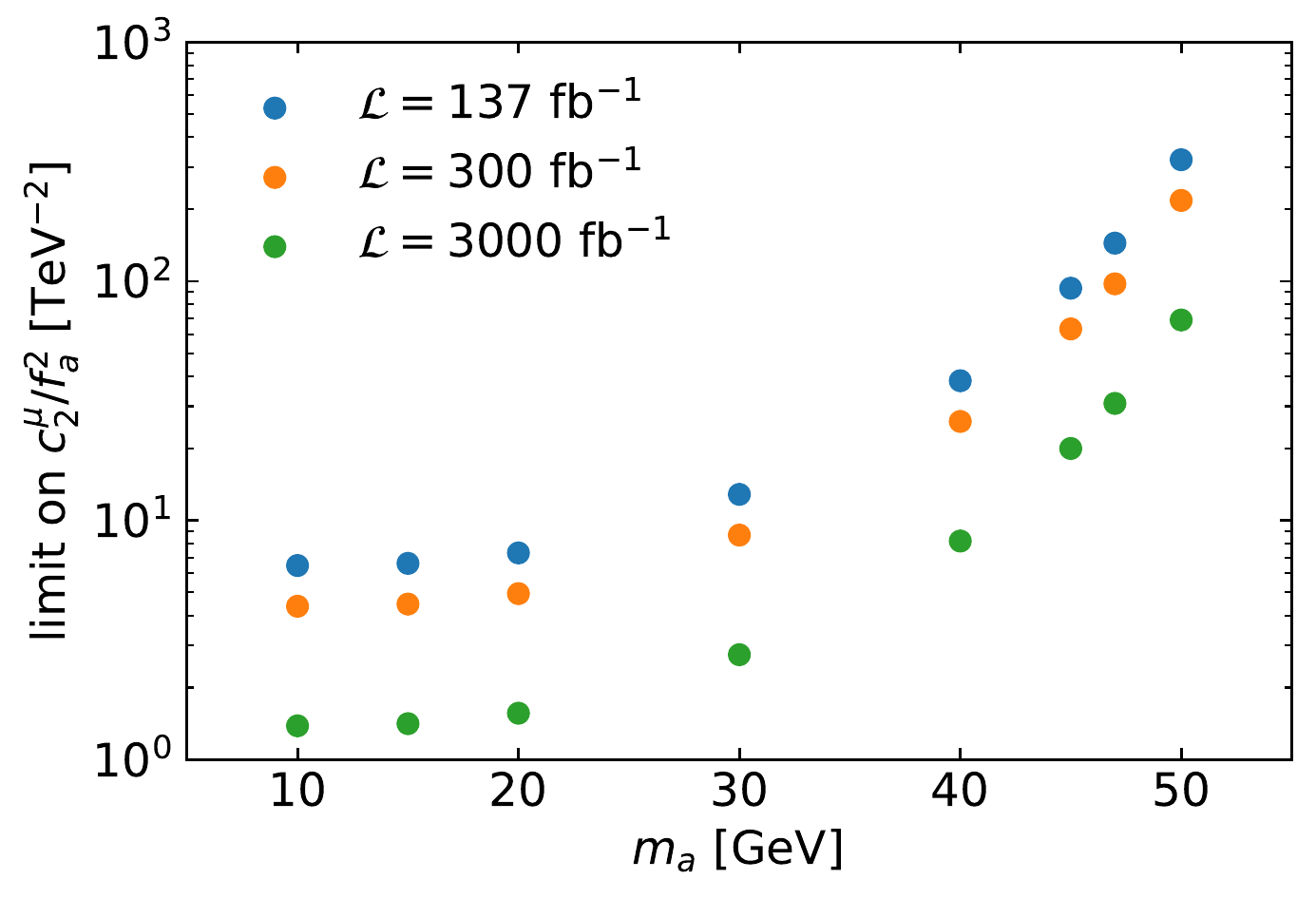} 
		\caption{Top: $m_{12}$ distribution in the $h \to a a \mu \mu \to  6 \mu$ analysis for different ALP masses at $\mathcal{L} = 137\ifb$.  Bottom: Upper $95\,\%$ CLs limits on $c_2^\mu/f_a$ for different luminosities.}
		\label{fig:6mu_limits}
\end{figure}

The ALP coupling $c_2^\mu$ induces a Higgs decay to six muons via 
the $h \to aa\mu \mu$ channel, where the ALP subsequently decays to muons. 

Several ATLAS and CMS searches are sensitive to $6 \mu$ final states. 
However, as discussed in Sec.~\ref{sec:theory}, even in background-free regions testing $\mathcal{O}
(1)$ $c_2^\mu$~couplings for $f_a=1\tev$ requires a signal acceptance of at least $10\%$.  
Analyses that require at least four muons are either insensitive, as a result of the larger background contributions~\cite{CMS:2019lwf}, or they require cuts on the lepton transverse momenta and sufficient energy, which render their signal acceptance below $10\%$~\cite{ATLAS:2018rns,CMS:2019hsm}. 

Analyses targeting a six-muon final state are currently focusing on specific resonances producing di-muon pairs.  
Ref.~\cite{CMS:2020hjs} explores the production of three $Z$ bosons in final states with six leptons.  
Since the leptons from $Z$~boson decays typically have relatively high transverse momenta, the analysis selects events in which the scalar sum of the $p_T$ of all leptons is above $250$ GeV, where only about $2\%$ of our signal events survive.
Ref.~\cite{CMS:2021qsn} concentrates on six-muon events ensuing from the production of three $J/\Psi$ mesons. 
The cut on the invariant mass of each opposite-sign
pair of muons, $m_{\mu\mu} \in [2.9,3.3]$ GeV, reduces our signal to insignificant levels 
(for the range of $m_a$ of interest). 
On the positive side, the search provides evidence that
the high-level trigger, together with mild off-line data requirements, can efficiently select multi-muon
events with muon transverse momenta as low as $6\gev$. 
Our following analysis relies on this fact.

We define our $h \to 6 \mu$ signal region through the minimal cuts:
\begin{equation}
p_T^\mu > 5 \gev \, , \quad 
m_{6 \mu}  \in [110, \, 140] \gev
\,.
\end{equation}
Potential backgrounds from low-mass $J/\psi$ or $\Upsilon$ resonances decaying to two muons are suppressed by the additional cut on the six muon invariant mass. At $\sqrt{s} = 13$~TeV, the SM cross section for $6\mu$~production (before requiring $m_{6 \mu}  \in [110, \, 140] \gev$)  is only 
$0.24~\text{ab}$. Therefore, we can safely neglect the SM background for LHC luminosities 
below $3\, \text{ab}^{-1}$.

We estimate the experimental efficiency of the $6\mu$ final state based on the $4\mu$ analysis in Ref.~\cite{CMS:2021pcy}.  We approximate the total efficiency per lepton as $93\%$ and exponentiate this with six.  The total deduced efficiency is $64\, \% $. 
To determine the ALP mass from the $6\mu$ analysis, we assign opposite-sign lepton pairs such that the difference $\Delta m = m_{12} - m_{34}$ between the invariant mass of two of these pairs is minimal, with $m_{12} > m_{34}$.  The resulting $m_{12}$ distribution for different ALP masses, which peaks at $m_{12} \sim m_a$, is shown in the top panel of Fig.~\ref{fig:6mu_limits}.

We use the $m_{12}$ distribution to constrain $c_2^\mu/f_a^2$ and find the following $95\,\%$ CLs limits at $\mathcal{L} = 137 \ifb$:
\begin{equation}
 \frac{c_2^\mu}{f_a^2} <
 \begin{aligned}
 \hspace*{0.8ex} 6.5/\text{TeV}^2 \quad  \text{for }m_s = 10\gev \\  
320/\text{TeV}^2 \quad \text{for }m_s = 50\gev
 \end{aligned}
  \,  .
\end{equation}
In the bottom panel of Fig.~\ref{fig:6mu_limits}, we show the limits on $c_2^\mu/f_a^2$ for different ALP masses and luminosities. 
For high ALP masses, the limits on $c_2^\mu$ become very weak due to the significant phase-space suppression.  
Since this channel is background free, the limits scale with $\mathcal{L}^{-1/2}$.

\subsection{Four muon two jet final state}
\label{sec:analysis_4mu2q}
\begin{figure}[thb]
		\centering
		\includegraphics[width=.2\textwidth]{figs/decay_4mu2q.pdf} \quad
		\includegraphics[width=.2\textwidth]{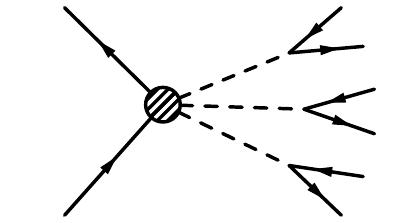}
		\caption{Feynman diagrams for $ p p  \to 4 \mu 2 j$ involving a five-point interaction of two quarks, two ALPs and the Higgs.}
		\label{fig:feyn_4mu2q}
\end{figure}

\begin{figure}[thb]
		\centering
		\includegraphics[width=.45\textwidth]{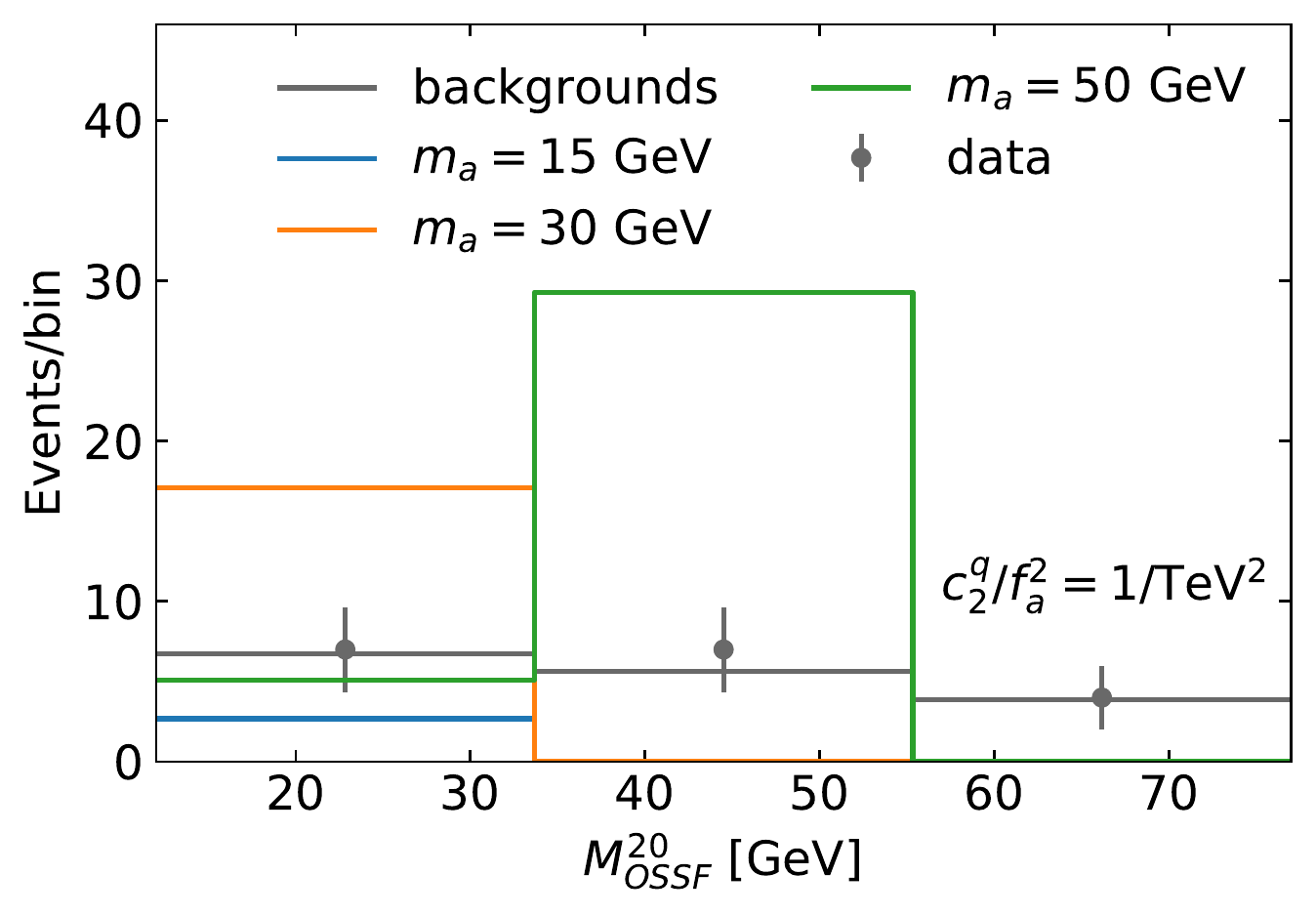} \\
		\includegraphics[width=.45\textwidth]{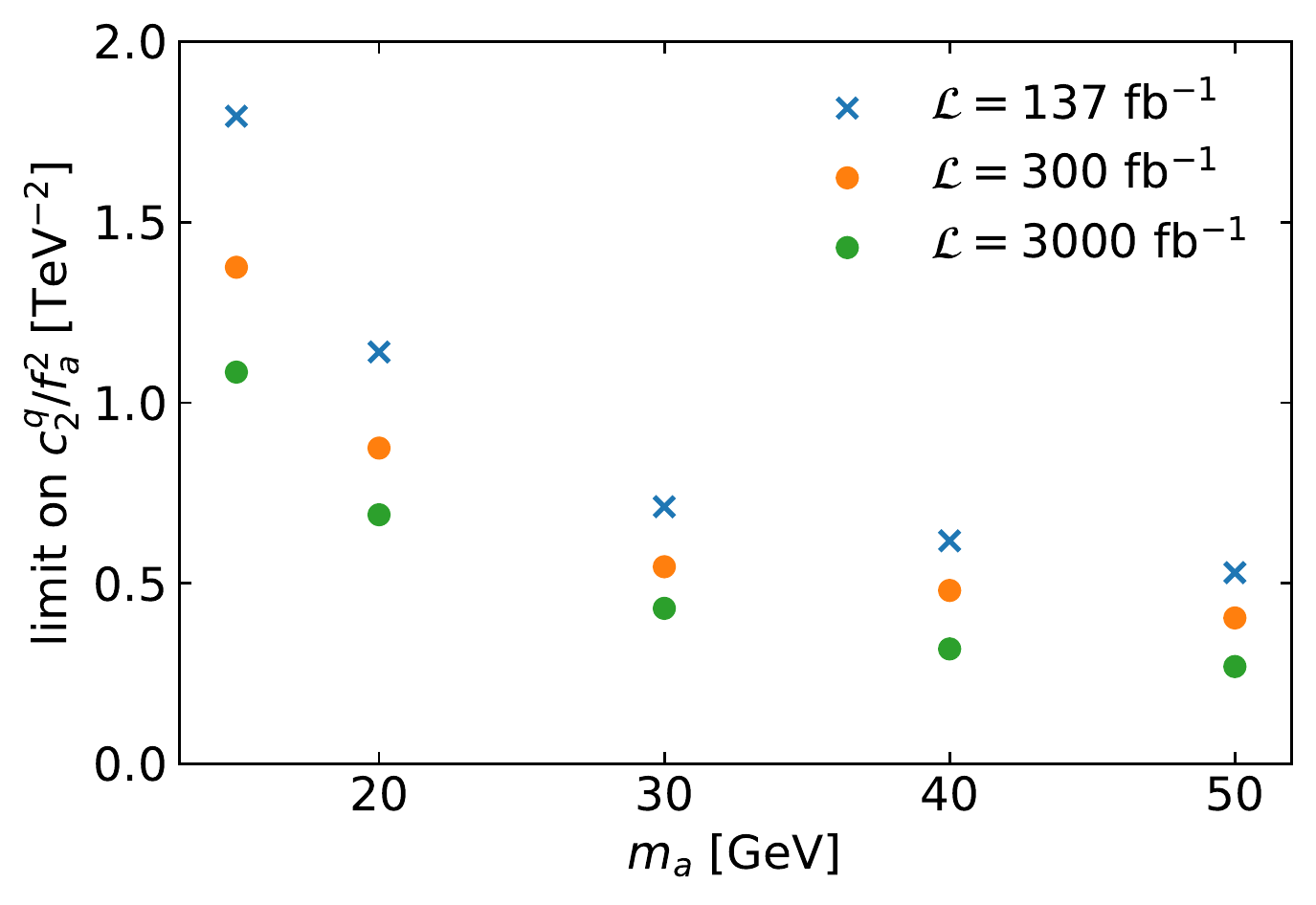} 
		\caption{Top: $M_\text{OSSF}^{20}$ distribution in the $p p \to h aa \to 4 \mu 2j$ analysis for backgrounds,  CMS data~\cite{CMS:2019lwf} and ALP signal for different masses ($\mathcal{L} = 137 \ifb$).  Bottom: Upper $95\,\%$ CLs limits on $c_2^u/f_a$ at different luminosities.  Observed limits are marked with an ``x''.}
		\label{fig:4mu2q_limits}
\end{figure}

Finally, we consider a five-point interaction of two ALPs with the Higgs boson and two quarks. The corresponding Wilson coefficient $c_2^q$ denotes a universal coupling to both up-type and down-type second-generation quarks. 
Assuming that the ALPs decay to muons, a non-zero $c_2^q$ coupling produces a $4 \mu 2 j$ signature at the LHC via two different channels shown in Fig.~\ref{fig:feyn_4mu2q}: First, the signature can result from a Higgs decaying into two ALPs and two jets $ p p \to h \to 2a 2 j$. And second, the coupling to second-generation quarks on the production side induces the $ p p \to h a a$ (with $h \to j j$) channel.  
The Higgs decay channel $h \to aa j j $ suffers from a large phase-space suppression and is only open for ALP masses $m_a \lesssim 36 \gev$ when we require the two jets to have transverse momenta greater than $25\gev$. Since the $h$ and $haa$ production channels do not interfere, we can distinguish their respective cross sections and find that the $haa$ cross section is 250 (24$\times 10^{6}$) times larger than the $h$ channel cross section for an ALP with mass $m_a = 15 \gev$ ($m_a = 30 \gev$). 
Focussing on the Higgs decay region only, for instance
by cutting on the invariant mass of the 4-muon-2-jets system would thus significantly weaken the limits on $c_2^q/f_a^2$.  
Here, we investigate a generic $4 \mu 2j$ ALP signature and present an analysis focussing on Higgs decays only in Appendix~\ref{app:4mu2j}.

The CMS search for beyond the SM phenomena in multi-lepton final states~\cite{CMS:2019lwf} provides a signal region with good sensitivity to our signal process
\footnote{Most experimental searches testing exotic signals in $4\mu2j$ final states are insensitive to our signal since they apply relatively strong cuts on the lepton transverse momenta or on their effective mass~\cite{ATLAS:2018rns,CMS:2019hsm}. 
The CMS search for pseudo-scalars decaying to a $Z$ boson and a Higgs, in events with four leptons and
two jets~\cite{CMS:2019kca} is insensitive to our ALP signal since it requires at least one electron.}. 
Since the Higgs boson predominantly decays to bottom quarks, the final-state jets in our $h a a$ analysis are likely to be $b$~jets. Therefore, we can recast the $4L(\mu\mu) \, 1B$ signal region of the CMS search, which requires four leptons and at least one $b$-tagged jet. To suppress the $ZZ+$jets background, the invariant mass of OSSF lepton pairs must not be within $15 \gev$ of the $Z$~boson mass:
\begin{equation}
\begin{split}
N_\ell \geq 4 \, , \quad
N_\text{OSSF}\geq &1 \, , \quad
p_T^{\ell} > 10 \gev \, ,\\
|\eta^{\mu (e)}| < 2.4 (2.5),  \quad 
&p_T^{\mu_1 (e_1)}  > 26 (35) \gev \, , \\
| m_{\ell^+\ell^-} - m_Z | < 15 \gev  \, , \quad
&M_\text{OSSF}^{20} (\mu \mu) \in [12, \, 77] \gev , \\
p_T^j > 30 \gev \, , \quad
&|\eta^j| < 2.1 \, , \quad
N_b \geq 1 \, , \\
\Delta R (\ell_i , \ell_j/j ) > 0.4 \, , \quad &
m_{\ell\ell, SF} > 12 \gev
\, . 
\end{split}
\label{eq:cuts_4mu2q}
\end{equation}
The variable $\mOSSF$ denotes the invariant mass of the two OSSF leptons with an invariant mass closest to the target mass of $20\gev$. 
Since the same target mass of $20\gev$ is applied for the pairing of the OSSF leptons, high-mass ALPs can contribute to more than one bin of the $\mOSSF$ distribution when leptons from different ALP decays are paired. 
We show the distribution of $\mOSSF$ for the SM backgrounds and different mass ALP signals in the top panel of Fig.~\ref{fig:4mu2q_limits}.  
The background, which is dominated by $tt\bar Z$ and $ZZ$ production, is directly taken from the \texttt{HepData}~\cite{Maguire:2017ypu} entry for Fig.~10e of Ref.~\cite{CMS:2019lwf}. We use the same reference for the systematic uncertainty on the background distribution, which is approximately $25\%$ in each bin. 
We have estimated the detector efficiency to be $64\,\%$,  analogous to the $6\mu$ final state.
To ensure that the EFT description is valid, we require the invariant mass of all selected muons and jets to be smaller than $1\tev$.

In Fig.~\ref{fig:4mu2q_limits},  we present the limits on $c_2^q/f_a^2$ as a function of the ALP mass for different luminosities.  The limits for low $m_a$ are weaker as a result of the decreased signal acceptance, in particular for the muon transverse momentum cuts. 
At $\mathcal{L} = 137 \ifb$ luminosity the observed (expected) $95\%$ CLs limits are:
\begin{equation}
 \frac{c_2^q}{f_a^2} <
 \begin{aligned}
 1.8 \, (1.8)/\text{TeV}^2 \quad  \text{for }m_s = 15\gev \\  
0.5 \, (0.5)/\text{TeV}^2 \quad \text{for }m_s = 50\gev
 \end{aligned}
  \,  .
\end{equation}

\section{Conclusions}
\label{sec:conclusions}
This paper investigates search strategies for a light pseudo-scalar particle in the $m_a \in [10, \, 50] \gev$ range coupling to the Higgs boson and fermions. These interactions induce rare new decays of the Higgs boson. We have focussed on second-generation leptons and have considered the possibility that the pseudo-scalar decays into two muons or two photons. 

The expected upper limits on the effective four-point interactions at $3\iab$ luminosity and assuming $f_a = 1 \tev$ are of the order of $c_1^\mu \lesssim 0.01$ for the $h \to 4 \mu$ analysis and $c_1^\mu \lesssim 0.03$ for the $h\to 2 \mu 2 \gamma$ analysis.  
Couplings inducing a five-particle interaction are less constrained and have a stronger mass dependence. We find $c_2^\mu \lesssim 1-80$ for the $h \to 6\mu$ channel and $c_2^q \lesssim 2.5-6.0$ for the $h \to 4\mu 2j$ channel (see the Appendix~\ref{app:4mu2j} for this later one), depending on the ALP mass. 
In terms of Higgs branching ratios these numbers correspond to $1.3 \times10^{-5}$, $2.0 \times10^{-6}$, $3.0 \times10^{-9}$ and $5.4 \times10^{-9}$ respectively. 

For $h \overline{q}q a a$ couplings, it turns out that despite the PDF suppression, the channel $p p \to a a h$ is by far more constraining than the Higgs decay $p p \to h \to q q a a$.  
Using the $pp \to a a h \to 4 \mu 2j$ channel, the upper limits on $c_2^q$ can be reduced to $c_2^q \lesssim 0.2-1$ at $3\iab$ luminosity and assuming $f_a = 1 \tev$.

There are four channels involving contact interactions of the Higgs with two fermions and ALPs with ALP decays to muons and photons which we did not consider here:
\begin{itemize}
\item $h\to a qq$, $a \to \mu \mu$ ($2\mu 2j$) or $p p \to h a \to b \bar b  \mu \mu$\,,
\item $h\to a qq$, $a \to \gamma\gamma$ ($2\gamma 2j$) or $p p \to h a \to b \bar b \gamma\gamma$\,,
\item $h \to aa\mu \mu$, $a\to \gamma\gamma$ ($4\gamma 2\mu$)\,,
\item $h \to aa q q $, $a\to \gamma\gamma$ ($4\gamma 2q$) or $p p \to h a a \to b \bar b +4 \gamma$\,.
\end{itemize}
We have neglected these channels due to the poorer reconstruction efficiency of low-energy photons compared to muons and the extensive backgrounds for events with two muons/photons and jets.

Dedicated searches for events with two leptons and more than two photons currently do not exist. The most promising search for a $2\mu4\gamma$ signature would thus still be the $Z\gamma\gamma$ search in Ref.~\cite{CMS:2021jji}, which already proved to have too stringent requirements on the photons for the $2\mu2\gamma$ signature studied in Sec.~\ref{sec:analysis_2mu2a}.

\section*{Acknowledgements} 
This work has been supported by the Newton International Fellowship Alumni follow-on funding 2020/21 under grant number AL211013/4.
A.B.~gratefully acknowledges support from the Alexander-von-Humboldt foundation as a Feodor Lynen Fellow.
MC is supported by the Spanish MINECO under the Ram\'on y Cajal programme as well as by the SRA under grant number PID2019-106087GB-C21/C22 (10.13039/501100011033), and by the Junta de Andaluc\'ia grants FQM 101, A-FQM-211-UGR18 and P18-FR-4314 (FEDER).

\appendix
\section{Four muon and two jet final state in Higgs decays}
\label{app:4mu2j}

\begin{figure}[thb]
		\centering
		\includegraphics[width=.45\textwidth]{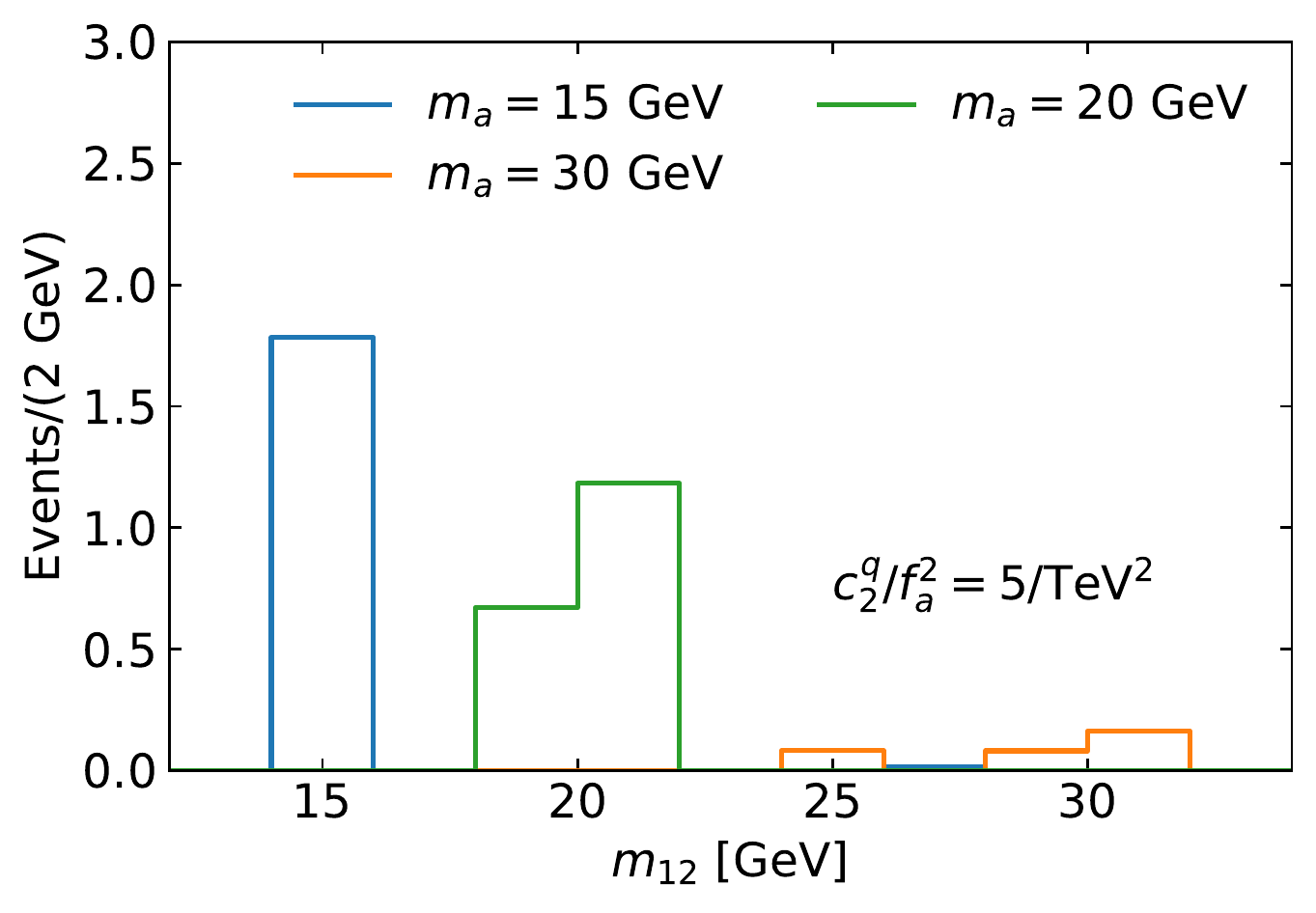} 
		\includegraphics[width=.45\textwidth]{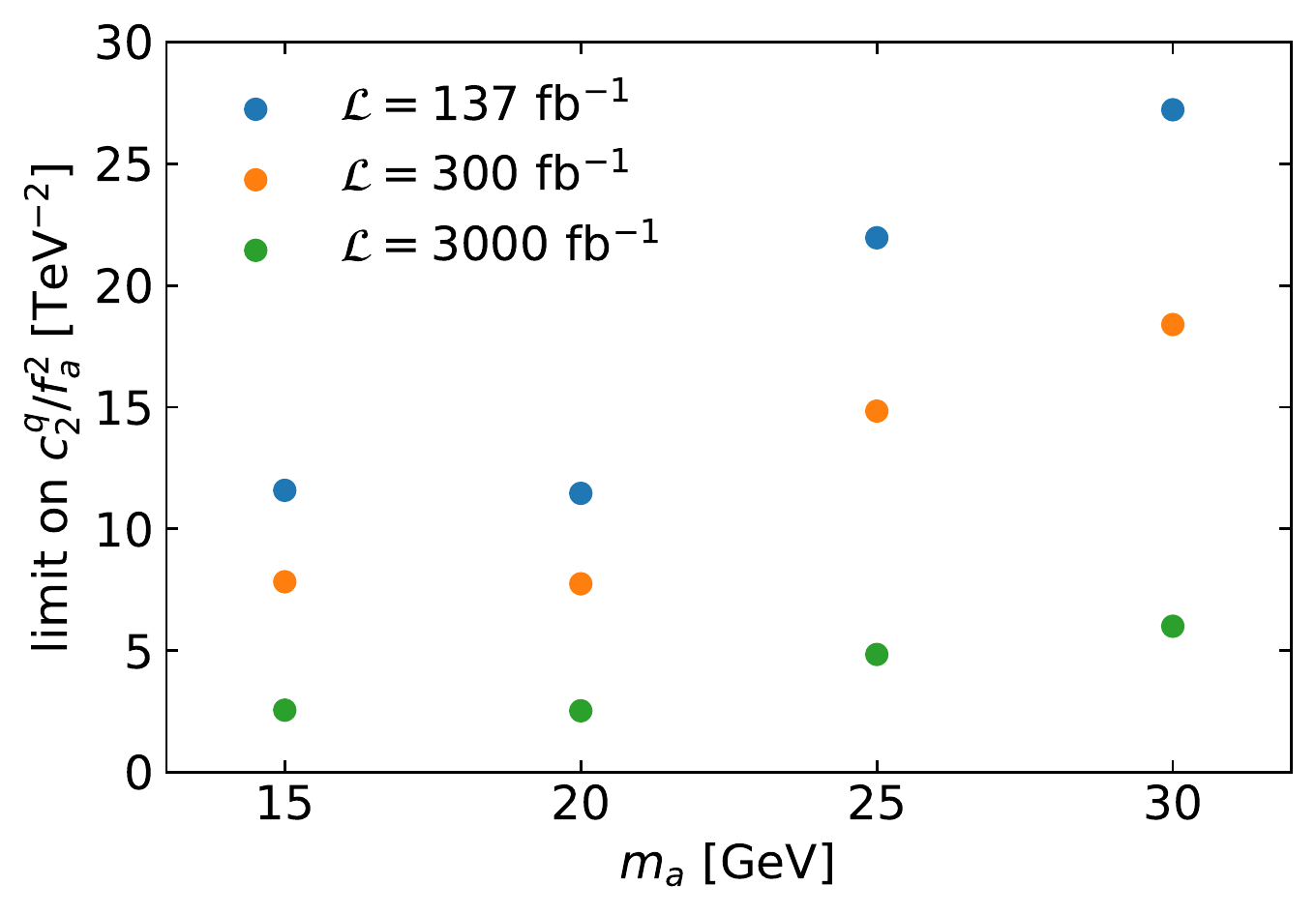} 
		\caption{Top: Invariant mass distribution of the heavier opposite-sign muon pair $m_{12}$ in the $h \to aa jj $ analysis at $\mathcal{L} = 137\ifb$.  Bottom: Upper $95\,\%$ CLs limits on $c_2^q/f_a^2$.}
		\label{fig:4mu2q_limits_hdecay}
\end{figure}

For completeness, we present an analysis for Higgs decays to two ALPs and two jets induced by a non-zero $c_2^q$.  We select the Higgs decay region by requiring that the invariant mass of the four muons and two jets is within $25\gev$ of the Higgs mass.  
While this selection ensures a better validity of the EFT approach,  it also significantly decreases the signal cross section as discussed in Section~\ref{sec:analysis_4mu2q}. 

As our signal is characterized by two resonant $a \to \mu \mu$ decays, we assign the two opposite-sign muon pairs such that the difference between their invariant mass,  $(m_{12}-m_{34})/(m_{12}+m_{34})$, is minimal and require $\Delta m = |m_{12} - m_{34}| < 4 \gev $.
The total signal selection is
\begin{equation}
\begin{split}
N_\mu \geq 4 \, , \quad
N_\text{OSSF}\geq 2 \, , \quad
&p_T^{\mu} > 5 \gev \, , \quad 
m_{\mu\mu} > 12 \gev ,\\
p_T^j > 25 \gev \, , \quad
&|\eta^j| < 4.5 \, , \quad
N_j \geq 2 
\\
| m_{\mu^+\mu^-} - m_Z | < 15 \gev  \, , \quad
&\Delta m = |m_{12} - m_{34}| < 4 \gev \\
m_{4\mu2j} \in [100&, \, 150] \gev
\, . 
\end{split}
\label{eq:cuts_4mu2q_hdecay}
\end{equation}

The dominant backgrounds from $ZZ + \text{jets}, \, Z \to 2 \mu$ production, $Z + \text{jets}, \, Z \to 4 \mu$ production and $h + \text{jets}, \, h \to 4 \mu$ production are significantly suppressed by the cuts on $m_{4\mu2j}$ and $\Delta m$.  After the selection cuts we expect less than one background event at the HL-LHC and we therefore consider our signal region as background free.  

To set limits we use the invariant mass pair of the heavier opposite-sign muon pair $m_{12}$;  see the upper panel of Fig.~\ref{fig:4mu2q_limits_hdecay}. We find the following upper CLs limits on $c_2^2/f_a^2$ at $\mathcal{L}=137\ifb$:
\begin{equation}
 \frac{c_2^q}{f_a^2} <
 \begin{aligned}
12 /\text{TeV}^2 \quad  \text{for }m_s = 15\gev \\  
27/\text{TeV}^2 \quad \text{for }m_s = 50\gev
 \end{aligned}
  \,  .
\end{equation}
In the bottom panel of Fig.~\ref{fig:4mu2q_limits_hdecay},  we present the  $95\%$ CLs limits on $c_2^q/f_a^2$ at different ALP masses and luminosities.  Comparing these limits to those shown in Fig.~\ref{fig:4mu2q_limits} for a generic $4\mu 2j$ signature analysis, we find that the limits obtained by focussing on Higgs decays only are worse by a factor $6- 40$ depending on the ALP mass.

\bibliographystyle{apsrev4-1}
\bibliography{Refs}

\end{document}